\documentclass[a4paper,11pt]{article}
\pdfoutput=1
\usepackage{jheppub}
\usepackage[T1]{fontenc}
\usepackage{graphicx,bm,amsmath,amssymb,mathtools,booktabs,placeins}
\allowdisplaybreaks
\newcommand{\dd}{\mathrm{d}}
\newcommand{\ii}{\mathrm{i}}
\newcommand{\Tr}{\mathrm{Tr}}
\newcommand{\bperp}{\bm b_\perp}

\newcommand{\Dperp}{\bm\Delta_\perp}
\newcommand{\kperp}{\bm k_\perp}
\newcommand{\calM}{\mathcal M}
\newcommand{\Dmod}{D_g^{\rm mod}}
\newcommand{\Dtmod}{\widetilde D_g^{\rm mod}}
\newcommand{\Mmod}{\mathcal M_{g,\rm mod}}
\newcommand{\pmodel}{p_g^{\rm mod}}
\newcommand{\smodel}{s_g^{\rm mod}}
\newcommand{\stf}[1]{\left[#1\right]_{\rm STF}}
\begin{document}
\title{Gluon stress and the D-term: small-\(x\) operator constraints and finite-range models}
\author[a]{Lei Wang}
\affiliation[a]{Key Laboratory of Quark and Lepton Physics (MOE)
 and Institute of Particle Physics, Central China Normal University,
 Wuhan, Hubei 430079, China}
\emailAdd{leiwang@ccnu.edu.cn}
\abstract{The leading small-x dipole relation for the gluon momentum form
factor motivates an analysis of the additional fields needed to describe
transverse stress. The spin-zero energy-momentum-tensor projector isolates
a mixed leading and subleading field-strength product absent from a
leading-plus-field background. We construct commuting, externally sourced
gauge fields with a specified screened response and calculate their
normalized stress functional. A weighted-moment identity fixes its forward
sign and bounds its magnitude for sufficiently regular radial profiles,
including profiles with sign-changing pointwise stress. For a Gaussian
profile, a positive-weight integral representation yields an explicit
error bound for the local approximation over a finite momentum-transfer
window and constrains the normalized curvature independently of the
response range. Coordinate-space stress relations provide regular
pressure and shear benchmarks. A separate Coulombic color-singlet
calculation compares filtered and unfiltered mixed fields in the same
target: its point-source limit has linear range sensitivity, while a
finite ultraviolet regulator changes the small-range behavior. A correlated
refit of author-supplied pion form factors separates overall fit quality
from range sensitivity. The full-window optimum lies at the infinite-range
shape boundary, while the forward extrapolation depends on the fitted model.
These results quantify approximation errors and inference limits within
specified stress models. Their connection to the physical small-x gluon
D-term requires renormalized QCD matching.}
\maketitle
\raggedbottom

\section{Introduction}
\label{sec:intro}
Gravitational form factors (GFFs) parameterize hadronic matrix elements of
the energy-momentum tensor (EMT). Their relation to moments of generalized
parton distributions (GPDs) connects this local operator to hard exclusive
processes \cite{Ji:1996ek,Ji:1996nm,Radyushkin:1997ki,Collins:1996fb};
the broader GPD framework is reviewed in Refs.~\cite{Diehl:2003ny,Belitsky:2005qn}.
The D-term completes the off-forward information carried by these moments
\cite{Polyakov:1999gs} and admits a mechanical interpretation through
spatial stress \cite{Polyakov:2002yz,Polyakov:2018zvc,arXiv180610591}.
Light-front (LF) densities give a two-dimensional realization of that
interpretation \cite{arXiv181009837,Freese:2021czn}. The distinction between
the EMT form factors and a chosen spatial-density prescription is central
to this interpretation \cite{Freese:2021mzg,Lorce:2025oot}.

Lattice QCD now constrains gluon GFFs and mechanical distributions in the
nucleon and pion \cite{Shanahan:2018pib,Shanahan:2018nnv}, and addresses
both spin dependence and quark--gluon separation
\cite{Pefkou:2021fni,Hackett:2023nkr,Hackett:2023rif}.
Near-threshold quarkonium measurements provide a complementary experimental
route \cite{Duran:2022xag}, whose relation to gluonic GFFs involves
process-dependent theoretical input
\cite{Hatta:2018ina,Guo:2021ibg,Guo:2025jiz}.
These developments motivate separating a physical EMT matrix element from
an auxiliary stress model before comparing either with phenomenology.
The present analysis concerns a spin-zero target.

At high energy, the leading color field is represented by lightlike Wilson
lines. The classical high-density source picture was developed for large
nuclei \cite{McLerran:1993ni,McLerran:1993ka,McLerran:1994vd}, including
its intrinsic small-x gluon distribution \cite{Jalilian-Marian:1996mkd}.
Its quantum evolution is organized by the Wilson renormalization group
\cite{Jalilian-Marian:1997jhx,Jalilian-Marian:1997ubg} and the color-glass
functional evolution equation \cite{Iancu:2000hn,Iancu:2001ad}.
The associated Wilson-line hierarchy and its evolution have equivalent
operator and functional formulations
\cite{Balitsky:1995ub,Weigert:2000gi,Mueller:2001uk}.
In the dipole picture \cite{Mueller:1994jq}, large-\(N_c\), large-nucleus closure gives
the nonlinear dipole equation \cite{Kovchegov:1999yj,Kovchegov:1999ua}.
Reviews of the Color Glass Condensate (CGC) and its evolution are given in
Refs.~\cite{Weigert:2005us,Gelis:2010nm}.

Even at eikonal accuracy, gluon distributions retain their Wilson-line
structure: the dipole and Weizs\"acker--Williams distributions enter
different hard processes \cite{Dominguez:2010xd,Dominguez:2011wm}, while
higher correlators and polarization carry further information
\cite{Dominguez:2011gc,Metz:2011wb}.
Classifications of quark correlation functions organize the relations among
generalized transverse momentum dependent distributions (GTMDs), GPDs and
transverse momentum distributions \cite{Meissner:2009ww}.
Gluon small-x Wigner representations and diffractive observables resolve
correlations between transverse position and momentum
\cite{Hagiwara:2016kam,Hatta:2016dxp,Hatta:2017cte}.
Hagiwara, Tong, and Xiao derived a small-x relation between the gluon
momentum GFF and the transverse Laplacian of the dipole amplitude
\cite{Hagiwara:2024aoq}. They also identified sub-eikonal corrections as a
route toward other GFFs. Our analysis builds on this distinction between
leading dipole information and the additional field content of stress.

The operator expansions for such corrections depend on the process and
the high-energy expansion parameter
\cite{Altinoluk:2014oxa,Altinoluk:2015gia,Chirilli:2018kkw,Chirilli:2021lif}.
Finite target extent and dynamical background fields generate distinct
next-to-eikonal contributions
\cite{Altinoluk:2015xuy,Altinoluk:2020oyd,Altinoluk:2021lvu}.
Polarized Wilson lines provide an explicit example of information beyond
the unpolarized eikonal ensemble \cite{Kovchegov:2015pbl,Kovchegov:2018znm}.
For the D-term, two questions must be kept separate: which local EMT
component carries the stress, and which additional target information is
required to represent that component at small x. The connection between
the local field identity and a nonlocal light-ray operator depends on the
operator's longitudinal moment and normalization.

Within the established EMT and LF density framework, we construct a sourced
background with a specified response kernel. A weighted-moment identity
fixes the forward sign and magnitude bound for screened radial profiles; for
a Gaussian profile, an explicit bound controls the error of the local
approximation over a finite transfer window. Coordinate-space stress
relations provide regular radial benchmarks. The physical GPD moment fixes
the renormalized normalization and gives an on-shell recoil subtraction;
its ERBL contribution identifies the independent information needed beyond
zero-skewness input. A weak-coupling color-singlet benchmark uses a specified
Coulombic wave function to compare a finite-range replacement with the
leading potential field. Finally, a correlated comparison of the Gaussian
shape family with author-supplied pion form factors tests the sensitivity to
the response length. The response kernel, its normalization, and the
longitudinal operator matching define the boundary between these model
results and a physical small-x gluon GFF.

We use natural units, \(g^{\mu\nu}=\mathrm{diag}(1,-1,-1,-1)\),
\(v^\pm=(v^0\pm v^3)/\sqrt2\), and
\(\Tr(t^at^b)=\delta^{ab}/2\). Transverse indices are Euclidean and run over
two dimensions. For any transverse tensor,
\(X_{\rm STF}^{ij}=(X^{ij}+X^{ji})/2-\delta^{ij}X^{kk}/2\).
The ultraviolet scale is \(\mu\). The separate ensemble label
\(Y=\ln(1/x_{\rm cut})\) labels the high-energy rapidity cutoff of the
ensemble, whereas the local EMT matrix element carries the scale \(\mu\).

\section{Physical form factor and transverse-density conventions}
\label{sec:projection}
\subsection{Spin-zero EMT projector}
Spin-zero off-forward quark correlation functions have a dedicated tensor
decomposition \cite{Meissner:2008ay}; pion-pair generalized distribution
amplitudes provide a crossed-channel route to pion GFFs
\cite{Kumano:2017lhr}. The D-term of elementary and composite scalar
systems also depends on their specified EMT and dynamics
\cite{Hudson:2017xug}. We fix the local EMT convention explicitly here.
One-particle states obey
\begin{equation}
 \langle p'|p\rangle=2p^+(2\pi)^3\delta(p'^+-p^+)
 \delta^{(2)}(\bm p'_\perp-\bm p_\perp).
 \label{eq:state-normalization}
\end{equation}
With \(p=P-\Delta/2\), \(p'=P+\Delta/2\), and \(t=\Delta^2\), we use
\begin{equation}
 \langle p'|T_g^{\mu\nu}(0;\mu)|p\rangle
 =2P^\mu P^\nu A_g(t;\mu)
 +\frac12(\Delta^\mu\Delta^\nu-g^{\mu\nu}t)D_g(t;\mu)
 +2m^2\bar C_g(t;\mu)g^{\mu\nu}.
 \label{eq:emt}
\end{equation}
Here \(m\) is the target mass. All three form factors are dimensionless.
Our \(D_g\) uses the \(C\) convention of Eq.~(2) in
Ref.~\cite{Hagiwara:2024aoq}. For the symmetric transverse stress,
the convention of Lorc\'e et al.~\cite{arXiv181009837} instead gives
\(D_g^{\rm here}=4C_g^{\rm there}\); their separately named \(D_g\)
belongs to an antisymmetric EMT structure. This conversion concerns the
tensor convention; the target considered here has spin zero. The operators and form factors
are renormalized in a common scheme, taken to be \(\overline{\rm MS}\),
with quark--gluon mixing implicit. We suppress \(\mu\) below when unambiguous.

The leading-order scale constraint on the singlet spin-two sector can be
made explicit. With \(D_\Sigma=\sum_f D_{q_f}\), fixed \(n_f\), and
\(\overline{\rm MS}\) conventions, the transverse STF projection obeys
\begin{equation}
 \frac{\dd}{\dd\ln\mu^2}\binom{D_\Sigma}{D_g}
 =\frac{\alpha_s}{2\pi}
 \begin{pmatrix}-4C_F/3&n_f/3\\4C_F/3&-n_f/3\end{pmatrix}
 \binom{D_\Sigma}{D_g}+O(\alpha_s^2).
 \label{eq:spin-two-rg}
\end{equation}
The column sums vanish, so the total D form factor is invariant under this
mixing. If a calculation has matched a pair of hard coefficients onto this
closed spin-two sector, their evolution is governed by the negative
transpose of the matrix in Eq.~\eqref{eq:spin-two-rg}. This supplies a
renormalization-group consistency condition for the matched coefficients.

This renormalization issue is distinct from the kinematic tensor projection.
The separate quark and gluon operators and their mass decompositions depend
on the renormalization prescription
\cite{Ji:1995sv,Lorce:2017xzd,Metz:2020vxd}, and their two-loop mixing has
been computed explicitly \cite{Panagopoulos:2020qcn}.
Quark--gluon trace-anomaly decompositions
\cite{Hatta:2018sqd,Tanaka:2018nae} constrain the twist-four
\(\bar C_{q,g}\) form factors \cite{Tanaka:2022wzy}.
The transverse D-sector and the isotropic stress of a nonconserved subsystem
remain distinct \cite{Lorce:2021xku}.

In the Drell--Yan frame, \(\Delta^+=\Delta^-=0\), \(P_\perp=0\),
and \(t=-\Dperp^2\). The relevant components are
\begin{align}
 \langle T_g^{++}\rangle&=2(P^+)^2 A_g(t),\label{eq:tpp-ag}\\
 \langle T_g^{ij}\rangle&=\frac{D_g(t)}2
 (\Delta^i\Delta^j-\delta^{ij}\Dperp^2)
 -2m^2\bar C_g(t)\delta^{ij}.\label{eq:tij}
\end{align}
Angle brackets in these two equations denote the matrix element in
Eq.~\eqref{eq:emt}. Defining
\begin{equation}
 Q^{ij}=\Delta^i\Delta^j-\tfrac12\delta^{ij}\Dperp^2,
 \qquad Q^{ij}Q^{ij}=\tfrac12(\Dperp^2)^2,
\end{equation}
gives, for nonzero transverse transfer,
\begin{equation}
 D_g(t)=\frac{4Q^{ij}\langle p'|T_g^{ij}|p\rangle}{(\Dperp^2)^2}.
 \label{eq:cg-projector}
\end{equation}
The forward value is defined by the limit of the projector as
\(\Dperp^2\to0\). The \(\bar C_g\) contribution drops out of this STF
projection. At nonzero skewness \(\xi=-\Delta^+/(2P^+)\), however,
\begin{equation}
 \langle T_g^{++}\rangle=2(P^+)^2[A_g(t)+\xi^2D_g(t)].
 \label{eq:skewness}
\end{equation}
The D-term enters \(T_g^{++}\) at nonzero skewness through the \(\xi^2D_g\)
term in Eq.~\eqref{eq:skewness}.
Nonzero-skewness small-x evolution of unpolarized GPDs and GTMDs has been
studied within the dipole framework \cite{Kovchegov:2025yyl}.
The skewness and real-part corrections used for exclusive amplitudes are
are process-specific ingredients \cite{Kovchegov:2026gwb}; the model
normalization introduced below is specified separately.

\subsection{Renormalized GPD moment and recoil subtraction}
\label{sec:gpd-moment}
The physical normalization is fixed through the established collinear
GPD operator, independently of a proposed response kernel
\cite{Ji:1996nm,Polyakov:1999gs,Diehl:2003ny,Belitsky:2005qn}.
For the spin-zero target, use the gluon convention
\(H_g(x,0,0;\mu)=xg(x;\mu)\) for \(x>0\), with even extension to
negative \(x\). With \(z^\mu=(0,z^-,\bm0_\perp)\), its operator
definition is
\begin{equation}
 H_g(x,\xi,t;\mu)=\frac1{P^+}\int\frac{\dd z^-}{2\pi}
 e^{\ii xP^+z^-}
 \left\langle p'\left|
 \sum_{i=1}^2 F_a^{+i}(-z/2)\,
 \mathcal W_{ab}[-z/2,z/2]F_b^{+i}(z/2)
 \right|p\right\rangle_{\!\mu},
 \label{eq:gpd-definition}
\end{equation}
where \(\mathcal W\) is a straight adjoint lightlike link and the
charge-even gluon operator is understood. Taking the complete moment in
the renormalized twist-two operator expansion gives
\begin{equation}
 \begin{aligned}
 M_g(\xi,t;\mu)&\equiv\frac12\int_{-1}^{1}\dd x\,H_g(x,\xi,t;\mu)
 =\int_0^1\dd x\,H_g(x,\xi,t;\mu)\\
 &=\frac{\langle T_g^{++}\rangle}{2(P^+)^2}
 =A_g(t;\mu)+\xi^2D_g(t;\mu).
 \end{aligned}
 \label{eq:gpd-renormalized-moment}
\end{equation}
The factor of one half in the signed integral follows from gluon evenness.
Fourier completeness supplies the factor \(1/(P^+)^2\), and the
coincident transverse field product is \(T_g^{++}\).
The moment and local operator must use the same renormalization scheme,
including singlet quark--gluon mixing. Their equality is the standard
renormalized GPD sum rule. It fixes the longitudinal normalization of the
full moment; a rapidity-restricted dipole average requires a separate
matching calculation.

Equation~\eqref{eq:gpd-renormalized-moment} yields an exact extraction of
\(D_g\) from two moments at the same invariant transfer:
\begin{equation}
 D_g(t;\mu)=\frac{M_g(\xi,t;\mu)-M_g(0,t;\mu)}{\xi^2},
 \qquad \xi\ne0.
 \label{eq:gpd-fixed-t-subtraction}
\end{equation}
Matching the transverse transfer instead would mix the \(A_g\) and
\(D_g\) contributions. To see this, set
\(\kappa=\bm\Delta_\perp^2\) and impose both external mass shells in
the symmetric frame. Then
\begin{equation}
 t(\xi,\kappa)=-\frac{\kappa+4m^2\xi^2}{1-\xi^2},\qquad
 \widehat M_g(\xi,\kappa)=M_g\bigl(\xi,t(\xi,\kappa)\bigr).
 \label{eq:on-shell-recoil}
\end{equation}
At fixed \(t<0\), the nonzero-skewness matrix element in
Eq.~\eqref{eq:gpd-fixed-t-subtraction} must therefore use
\(\kappa=-(1-\xi^2)t-4m^2\xi^2\geq0\), whereas the zero-skewness
one uses \(\kappa=-t\). The allowed range is
\(\xi^2\leq -t/(4m^2-t)\); at \(t=0\), fixed-\(t\) extraction
requires a limit from spacelike transfer.
Alternatively, differentiating along the physical fixed-\(\kappa\) path
gives
\begin{equation}
 \boxed{D_g(-\kappa;\mu)=
 \left.\frac12\frac{\partial^2\widehat M_g}{\partial\xi^2}
 \right|_{\xi=0,\kappa}
 +(4m^2+\kappa)A_g'(-\kappa;\mu).}
 \label{eq:recoil-corrected-D}
\end{equation}
Here \(A_g'\) denotes a derivative with respect to \(t\).
This subtraction follows from on-shell kinematics and the EMT
decomposition. For example,
the published pion monopole central fits in Ref.~\cite{Hackett:2023nkr},
Tables III--IV, with \(m_\pi=0.170\,\mathrm{GeV}\),
\(A_g(t)=0.546/[1-t/(1.129\,\mathrm{GeV})^2]\), and
\(D_g(t)=-0.596/[1-t/(0.677\,\mathrm{GeV})^2]\), give a correction
of \(0.0495\) at \(\kappa=0\). Omitting it would yield
\(-0.6455\) in place of \(-0.596\). At
\(\kappa=0.4583\,\mathrm{GeV}^2\), the corresponding values are
\(-0.4310\) and \(-0.2980\). These numbers illustrate the exact
kinematic correction using the published central fits.

\subsection{Independent ERBL information}
\label{sec:erbl-information}
The physical moment requires information beyond zero-skewness input.
The standard D-term freedom \cite{Polyakov:1999gs,Diehl:2003ny} makes
this explicit. For any chosen function \(d(t;\mu)\), consider
\begin{equation}
 \delta H_g(x,\xi,t;\mu)=\frac{15}{8}\,|\xi|\,d(t;\mu)
 \left(1-\frac{x^2}{\xi^2}\right)^2
 \theta(|x|<|\xi|),\qquad \xi\ne0,
 \label{eq:erbl-deformation}
\end{equation}
with zero value at \(\xi=0\). It preserves evenness in \(x\) and
\(\xi\), vanishes at the ERBL endpoints, and obeys
\begin{equation}
 \int_0^1\dd x\,x^{2k}\delta H_g
 =\frac{15\,\xi^{2k+2}}{(2k+1)(2k+3)(2k+5)}d(t;\mu),
 \qquad k=0,1,\ldots.
 \label{eq:erbl-polynomiality-check}
\end{equation}
Thus it supplies an allowed highest polynomial power in every even gluon
moment and shifts \(D_g(t;\mu)\) by \(d(t;\mu)\), while leaving
\(H_g(x,0,t;\mu)\) and \(A_g(t;\mu)\) unchanged. For any fixed
\(x\ne0\), this contribution disappears in a sufficiently small
neighborhood of \(\xi=0\). A pointwise small-skewness expansion at
fixed nonzero \(x\) therefore misses its contribution to the quadratic
coefficient of the integrated moment.

This construction illustrates the established GPD D-term ambiguity;
additional dynamics constrain which functions \(d\) can describe a physical
hadron. Support, symmetry, polynomiality, and zero-skewness leading-field
input leave this information independent. A predictive construction must
compute the finite-skewness moment or supply an independent ERBL input.
Fitting \(d\) to a lattice D form factor calibrates that input, so an
independent test requires a separate observable or data set.

\subsection{Fourier density and mass-rescaled stress}
Impact-parameter parton densities provide the established setting for a
transverse spatial interpretation at zero skewness
\cite{Burkardt:2000za,Burkardt:2002hr}.
Localized-state and convolution analyses specify how such densities are
related to internal hadron structure
\cite{Freese:2021mzg,Freese:2022fat}.
At fixed \(P^+\), define the physical STF density by the Fourier pair
\begin{align}
 \calM_g^{ij}(\bperp;\mu)
 &=\int\dd^2\Dperp\,e^{+\ii\Dperp\cdot\bperp}
 \langle p'|T_{g,\rm STF}^{ij}(0;\mu)|p\rangle,\label{eq:mij-def}\\
 \langle p'|T_{g,\rm STF}^{ij}(0;\mu)|p\rangle
 &=\int\frac{\dd^2\bperp}{(2\pi)^2}e^{-\ii\Dperp\cdot\bperp}
 \calM_g^{ij}(\bperp;\mu).\label{eq:mij-inverse}
\end{align}
This placement of \((2\pi)^2\) makes \(\calM_g^{ij}\) a rescaled
inverse transform of mass dimension four. The conventional light-front (LF)
density uses the reduced matrix element divided by \(2P^+\)
\cite{arXiv181009837}, so
\begin{equation}
 \mathsf T_{g,\rm LF,STF}^{ij}(\bperp;P^+,\mu)
 =\frac{\calM_g^{ij}(\bperp;\mu)}{2P^+(2\pi)^2}.
 \label{eq:lf-density}
\end{equation}
The even spin-zero tensor is insensitive to reversing the Fourier phase.
These definitions imply the exact physical identity
\begin{equation}
 D_g(t;\mu)=\frac{4Q^{ij}}{(\Dperp^2)^2}
 \int\frac{\dd^2\bperp}{(2\pi)^2}e^{-\ii\Dperp\cdot\bperp}
 \calM_g^{ij}(\bperp;\mu).
 \label{eq:master}
\end{equation}
This physical identity is defined by the local EMT at renormalization scale
\(\mu\).

Define \(\widetilde D_g(b)=\int\dd^2\Dperp\,
 e^{\ii\Dperp\cdot\bperp}D_g(-\Dperp^2)/(2\pi)^2\), and use the same
measure for \(\widetilde{\bar C}_g\). Before taking the STF part, the
mass-rescaled LF transverse density is
\begin{equation}
 \tau_g^{ij}\equiv\frac{P^+}{m}\mathsf T_{g,\rm LF}^{ij}
 =\frac{\delta^{ij}\nabla_b^2-\partial_b^i\partial_b^j}{4m}
 \widetilde D_g-m\delta^{ij}\widetilde{\bar C}_g.
 \label{eq:mass-rescaled-density}
\end{equation}
The D-sector coefficients in
\(\tau_{D,g}^{ij}=\delta^{ij}p_{D,g}+
 (\hat b^i\hat b^j-\delta^{ij}/2)s_{D,g}\) are
\begin{equation}
 p_{D,g}=\frac{\widetilde D_g''+\widetilde D_g'/b}{8m},
 \qquad
 s_{D,g}=-\frac{\widetilde D_g''-\widetilde D_g'/b}{4m}.
 \label{eq:physical-mechanical-def}
\end{equation}
Both have mass dimension three. The factor \(1/(2m)\) is an explicit
\(P^+/m\) rescaling of the two-dimensional LF density. The full gluon isotropic stress also contains
\(-m\widetilde{\bar C}_g\). The D-sector tensor is divergence-free by
construction, but the partial gluon EMT need not be conserved
\cite{Lorce:2021xku,Lorce:2025oot}.
The two-dimensional pressure and shear relations are part of the LF
stress framework \cite{arXiv181009837,Freese:2021czn}; the explicit mass
rescaling above states the convention in which we use them.

\section{Local stress and high-energy operator information}
\label{sec:operator}
The classical field convention for the gluon EMT is
\begin{equation}
 T_g^{\mu\nu}=-2\Tr(F^{\mu\alpha}F^\nu{}_{\alpha})
 +\tfrac12g^{\mu\nu}\Tr(F^{\alpha\beta}F_{\alpha\beta}).
\end{equation}
Quantum composite operators require the renormalization specified above.
Its local transverse STF part is
\begin{equation}
 T_{g,\rm STF}^{ij}=-2\Tr\stf{F^{i+}F^{j-}+F^{i-}F^{j+}-F^{ik}F^{jk}}.
 \label{eq:local-stress}
\end{equation}
In two transverse dimensions \(F^{ij}=\epsilon^{ij}B\), and
\(F^{ik}F^{jk}=\delta^{ij}B^2\), including for matrix-valued \(B\).
The last term therefore vanishes after the STF projection. A background with
only \(A^+\), with \(F^{i-}=F^{ij}=0\), has zero local STF stress even
though its \(T^{++}\) is nonzero. This result characterizes the specified
background approximation.
An independent recent analysis finds vanishing transverse stress in an
eikonal approximation for a gauge-invariant canonical gluon EMT
\cite{Uji:2026gravitational}. Its momentum-space construction uses a
different EMT and gauge-link prescription from the off-forward local kinetic
D-term studied here.

For a boost by \(\zeta\) in the plus direction, the reference-field counting
is \(F^{i+}\sim\zeta\), \(F^{i-}\sim\zeta^{-1}\), and
\(\dd z^-\sim\zeta^{-1}\). Consequently \(T^{++}\sim\zeta^2\)
and the mixed transverse stress is of order \(\zeta^0\).
We use \emph{sub-eikonal} to denote the additional field content.
The order in a scattering amplitude also depends on its prefactors and
longitudinal kernels. Geometric twist is a separate classification: the transverse STF
stress is a component of the same dimension-four, spin-two local operator
whose plus-plus moment appears in Eq.~\eqref{eq:gpd-renormalized-moment}.
At finite skewness, the standard twist-two GPD moment contains the physical
D-term, even though transverse stress is suppressed relative to \(T^{++}\)
under a boost.

The process dependence is concrete in next-to-eikonal DIS dijet
calculations \cite{Altinoluk:2022jkk,Agostini:2024xqs} and in their
back-to-back expansion involving twist-two and twist-three gluon TMDs
\cite{Altinoluk:2024zom}.
In particular, mixed transverse and boost-enhanced target-field correlators
arise beyond the standard leading-field input \cite{Agostini:2024xqs}.
Gaussian target models for decorated correlators have also been constructed
for these observables \cite{Agostini:2025vvx}. The weighted radial stress
functional studied here addresses a different quantity within this broader
operator framework.

For an off-forward coherent amplitude, the color-singlet dijet projection
must be made before taking the target matrix element. In the generalized
eikonal contribution of Ref.~\cite{Altinoluk:2022jkk}, its retained
longitudinal center coordinate supplies the longitudinal recoil constraint
upon projection between momentum eigenstates. By contrast, some displayed
explicit next-to-eikonal terms there are expanded at zero longitudinal
exchange. Their extension to a nonzero-recoil coherent amplitude requires
consistent power counting and a derivation of the off-forward operator.
In particular, the bidirectional longitudinal derivative in the
dynamical-target term is distinct from the total derivative of the dipole
and carries information beyond the external momentum transfer.

The helicity sector illustrates why extra field content requires its own
operator evolution. The polarized evolution framework
\cite{Kovchegov:2015pbl,Kovchegov:2016zex} led to helicity-dependent
extensions of JIMWLK and of the MV ensemble
\cite{Cougoulic:2019aja,Cougoulic:2020tbc}.
The revised quark--gluon evolution system and its large-\(N_c\) analytic solution
\cite{Cougoulic:2022gbk,Borden:2023ugd} also show the importance of the
operator basis. These results illustrate the construction of evolution
equations for operators with additional field content.

To identify the relevant Wilson-line insertion, let
\begin{equation}
 U(z_2^-,z_1^-;\bperp)=\mathcal P\exp\left[
 ig\int_{z_1^-}^{z_2^-}\dd z^- A^+(z^-,\bperp)\right],
\end{equation}
where the endpoints below have common \(z^+\) and transverse position.
There is a direct, regulated equal-separation relation for a finite link.
Write \(z_\pm^-=z^-\pm\lambda/2\) and define
\begin{align}
 \mathcal B^{ij}(\lambda;z)&=-2\Tr\stf{F^{i+}(z_+)
 U(z_+,z_-)F^{j-}(z_-)U(z_-,z_+)}\nonumber\\
 &\quad-2\Tr\stf{F^{i-}(z_+)
 U(z_+,z_-)F^{j+}(z_-)U(z_-,z_+)},\label{eq:finite-link-bilocal}\\
 \widetilde{\mathcal B}^{ij}(k;z)&=\int\dd\lambda\,
 e^{ik\lambda}\mathcal B^{ij}(\lambda;z).
\end{align}
For regulated classical fields, or as a formal bare operator identity with
a common regulator, Fourier completeness gives
\begin{equation}
 \int_{-\infty}^{\infty}\frac{\dd k}{2\pi}\,
 \widetilde{\mathcal B}^{ij}(k;z)
 =\mathcal B^{ij}(0;z)=T_{g,\rm STF}^{ij}(z).
 \label{eq:finite-link-full-moment}
\end{equation}
The equality uses the full signed momentum integral and the vanishing of the
transverse STF part of \(F^{ik}F^{jk}\). It identifies a gauge-invariant
finite-link bilocal whose coincident limit gives the local stress. Its
renormalized moment and its coefficient in a scattering process require
ultraviolet renormalization, operator mixing, and matching to a light-ray
operator basis. A restriction to a small-\(x\) window also requires an
independent constraint on the remainder of the spectrum.

In contrast, consider two insertions integrated separately along infinite
lines and introduce
\begin{align}
 U_\nabla^i&=\int\dd z^-\,U(+\infty,z^-)\,igF^{i+}(z^-)U(z^-,-\infty),\\
 U_R^i&=\int\dd z^-\,U(+\infty,z^-)\,igF^{i-}(z^-)U(z^-,-\infty).
\end{align}
The transverse position is the same in both lines. Products such as
\begin{equation}
 \mathcal O^{ij}_{\rm cand}
 =-\frac2{g^2}\left\langle
 \Tr\stf{U_\nabla^iU_R^{j\dagger}+U_R^iU_\nabla^{j\dagger}}
 \right\rangle_Y
 \label{eq:decorated-basis}
\end{equation}
close the endpoint color transformations. Distinct transverse positions
would require the corresponding transverse closure links.
Since \((igF_1)(igF_2)^\dagger=g^2F_1F_2\), the displayed coefficient
reproduces the relative local two-field sign. Each decorated line has dimension one and the
product contains two independent longitudinal integrations. A longitudinal
kernel, a partonic moment, and ultraviolet and rapidity subtractions are
required to match the product to a local EMT matrix element; the full basis
can contain additional operators.

The independent integrations in Eq.~\eqref{eq:decorated-basis} differ from
the coincidence enforced by a complete Mellin moment. A delta-function kernel
in the relative longitudinal coordinate defines a different operator; its
kernel requires information beyond gauge closure. The standard, renormalized
connection to the physical D-term is Eq.~\eqref{eq:gpd-renormalized-moment}. Applying it to a particular
high-energy operator requires a matching that retains the independent ERBL
information identified in Section~\ref{sec:erbl-information}.

Rapidity factorization and matching require their own calculation.
Gluon TMD evolution connects low and moderate x
\cite{Balitsky:2014wna,Balitsky:2015qba}, with separate higher-twist
corrections \cite{Balitsky:2017flc}; regulator dependence is studied
explicitly in DIS factorization \cite{Altinoluk:2025tms}.
The evolution of a sub-eikonal dipole gluon distribution with a single
chromoelectric insertion has also been studied at large \(N_c\)
\cite{Li:2026subeikonal}. The mixed-stress longitudinal moment considered
here involves a different field insertion.
Recent sub-eikonal DIS analyses connect process-specific light-ray
operators to quark distributions
\cite{Chirilli:2026vij,Chirilli:2026pkv}. The longitudinal stress moment
and target normalization in Eq.~\eqref{eq:decorated-basis} remain separate
inputs to its matching.

Likewise, the integrated local operator
\begin{equation}
 \mathcal S_g^{ij}(\bperp)=\int\dd z^-T_{g,\rm STF}^{ij}(z^-,\bperp)
 \label{eq:stress-operator}
\end{equation}
has mass dimension three, whereas \(\calM_g^{ij}\) has dimension four. Between fixed momentum
states, translation supplies only an external phase, and the unrestricted
integral at \(\Delta^+=0\) contains a longitudinal volume factor.
The finite density in Eq.~\eqref{eq:mij-def} therefore requires its own
normalization.

Determining the mixed product in Eq.~\eqref{eq:local-stress} from correlators
of the leading \(A^+\) field requires an additional dynamical relation.
The example below illustrates this distinction using externally
sourced fields; the D-terms realized by physical hadrons depend on their
underlying dynamics.

\section{A finite-range response model with external sources}
\label{sec:response}
\subsection{Sourced commuting fields}
Choose a Hermitian color generator \(H\) with \(\Tr H^2=1/2\), a smooth
real longitudinal function \(h(z^-)\) of finite extent or rapid decay, and
a smooth radial transverse profile \(q(b)\). At a reference boost define
\begin{equation}
 A^+=\kappa h(z^-)q(b)H,\qquad A^i=0,\qquad
 A^-=-\sigma\epsilon R_0\kappa h(z^-)q_\ell(b)H,
 \label{eq:background}
\end{equation}
where \(\epsilon,R_0>0\), \(\sigma=+1,-1,0\), and
\begin{equation}
 q_\ell=(1-\ell^2\nabla_b^2)^{-1}q,
 \qquad
 \widetilde q_\ell(\kperp)=\frac{\widetilde q(\kperp)}{1+\ell^2\kperp^2}.
 \label{eq:screened-kernel}
\end{equation}
We impose regularity at the origin and decay at infinity. In this model
\([q]=2\), \([h]=0\), \([\kappa]=-1\), and \([\ell]=-1\), with
brackets denoting mass dimension. The identification of \(q\) with a
saturation-scale shape is an ansatz; its normalization and response
coefficient are independent model inputs.

Dipole fits already require an impact-parameter profile
\cite{Kowalski:2003hm,Kowalski:2006hc}; proton-geometry and fluctuation
studies add information beyond a single average radius
\cite{Schlichting:2014ipa,Mantysaari:2016ykx}.
Long-distance tails and infrared control are separate issues in
impact-parameter-dependent evolution
\cite{Kovner:2002xa,Ferreiro:2002kv,Golec-Biernat:2003naj}.
These studies motivate keeping the profile and response range explicit.
Equation~\eqref{eq:screened-kernel} specifies the response kernel used here.
Similarly, elliptic gluon phase-space correlations
\cite{Zhou:2016rnt,Hagiwara:2017ofm} and the mixed-field STF stress studied
here share a rank-two transverse angular structure while involving distinct
field correlators.

All color commutators vanish. The field strengths obey
\begin{equation}
 F^{i-}(\kperp)=-\frac{\sigma\epsilon R_0}{1+\ell^2\kperp^2}
 F^{i+}(\kperp),\qquad F^{ij}=0.
\end{equation}
They follow from actual
potentials, so the Bianchi identity is automatic. Taking
\(J^\nu=\partial_\mu F^{\mu\nu}\) supplies a conserved external color
current and solves the sourced Yang--Mills equations in this commuting
sector; the explicit current is given in Appendix~\ref{app:source}.
The Wilson line built from \(A^+\), and the classical leading
\(T^{++}\), are independent of \(\sigma\), whereas the external current and
mixed transverse stress depend on it. The construction specifies an external
current; a bound-state description would also require source matter dynamics
and the conserved total hadron EMT.

Let \(H_2=\int\dd z^-h(z^-)^2\). Direct substitution into
Eq.~\eqref{eq:local-stress} gives
\begin{equation}
 \Sigma_Y^{ij}\equiv\langle\mathcal S_g^{ij}\rangle_{Y,\rm cl}
 =2\sigma\epsilon R_0\kappa^2H_2
 \left(\hat b^i\hat b^j-\tfrac12\delta^{ij}\right)q'(b)q_\ell'(b).
 \label{eq:source-stress}
\end{equation}
The classical average can include a common global color rotation, which
leaves the trace and the result invariant. The label \(Y\) is held fixed
in all numerical examples.
The local squared-gradient coefficient is positive for the chosen commuting
fields. In a general
ensemble, a positive covariance with radial and tangential eigenvalues
\(E_r,E_t\geq0\) has STF coefficient \(E_r-E_t\), whose sign is unrestricted.

\subsection{Model normalization and sign conditions}
We introduce a positive reference normalization \(\mathcal N_{\rm mod}\)
of dimension one and define
\begin{equation}
 \Mmod^{ij}=\mathcal N_{\rm mod}\Sigma_Y^{ij}
 =\left(\hat b^i\hat b^j-\tfrac12\delta^{ij}\right)\Mmod^{(2)}(b),
 \quad
 \Mmod^{(2)}=\sigma\Lambda_*q'q_\ell',
 \label{eq:model-normalization}
\end{equation}
where \(\Lambda_*=2\epsilon R_0\mathcal N_{\rm mod}\kappa^2H_2>0\)
has dimension minus two. This gives \([\Mmod^{ij}]=4\), as in the physical
Fourier convention. The normalized model functional is
\begin{equation}
 \Dmod(t;Y)\equiv\frac{4Q^{ij}}{(\Dperp^2)^2}
 \int\frac{\dd^2\bperp}{(2\pi)^2}e^{-\ii\Dperp\cdot\bperp}
 \Mmod^{ij}(Y,\bperp).
 \label{eq:model-projector}
\end{equation}
The parameters \(\mathcal N_{\rm mod}\) and \(\Lambda_*\) set the model
normalization. Identifying \(\Mmod\) and \(\Dmod\) with the physical
quantities \(\calM_g\) and \(D_g\), or with a partonic x-window
contribution, requires a normalized target state and the longitudinal
matching discussed in Section~\ref{sec:operator}.

For a radial profile the angular integral yields
\begin{align}
 \Dmod(-\Delta^2)&=-\frac1{\pi\Delta^2}
 \int_0^\infty b\,\dd b\,J_2(\Delta b)\Mmod^{(2)}(b),
 \qquad \Delta=|\Dperp|,\label{eq:bessel}\\
 \Dmod(0)&=-\frac1{8\pi}\int_0^\infty b^3\,\dd b\,\Mmod^{(2)}(b).
 \label{eq:d0}
\end{align}
The local-response limit \(q_\ell=q\) therefore gives a negative forward
functional for \(\sigma=+1\), provided the nonzero weighted moment is
finite. At finite \(\ell\), the sufficient condition is
\(q'(b)q_\ell'(b)\geq0\). Reversing \(\sigma\) reverses the functional by
linearity. Static abelian force-field calculations
provide a separate example of how field stress affects a D-term sign
\cite{Freese:2026hjy}. Section~\ref{sec:quarkonium} gives a separate
leading-potential singlet benchmark with a specified bound-state wave function.
Section~\ref{sec:finite} proves a stronger
integrated result for the screened kernel, without requiring pointwise
positivity of this product.

\section{Coordinate-space stresses and smooth local benchmarks}
\label{sec:mechanics}
Define \(\Dtmod,\pmodel,\smodel\) by replacing \(D_g\) with \(\Dmod\)
in the Fourier transform and Eq.~\eqref{eq:physical-mechanical-def}.
These mass-rescaled diagnostics describe the model D-sector; the isotropic
\(\bar C_g\) contribution is treated separately in Eq.~\eqref{eq:mass-rescaled-density}.
The following are the standard two-dimensional stress relations in our
Fourier convention. In particular, Eq.~\eqref{eq:pressure-normalization}
is the integrated radial equilibrium relation in Sec.~V.1.2 of
Ref.~\cite{arXiv181009837}, applied to the divergence-free D-sector.
That reference's Eqs.~(97), (108)--(111), with \(D=4C\), give the
LF normalization and pressure/shear conventions used here.
The same LF mechanical structures and radial equilibrium relation are
derived in Sec.~IV.2 of Ref.~\cite{Freese:2021czn}; our factors follow
the mass-rescaled convention in Eq.~\eqref{eq:mass-rescaled-density}:

\begin{align}
 \smodel(b)&=\frac{\Mmod^{(2)}(b)}{8\pi^2m},\label{eq:shear-normalization}\\
 \pmodel(b)&=\int_b^\infty\frac{\dd u}{u}\smodel(u)-\frac12\smodel(b),
 \label{eq:pressure-normalization}\\
 \Dtmod(b)&=-\frac1{4\pi^2}\int_b^\infty\dd u\,
 \frac{u^2-b^2}{u}\Mmod^{(2)}(u).
 \label{eq:density-direct}
\end{align}
The boundary conditions are decay of the density and its stress at infinity.
They remove the homogeneous constant and \(b^2\) solutions. The derivation
and convergence requirements are given in Appendix~\ref{app:transforms}.
These formulas avoid differentiating a finite-bandwidth inverse Hankel
transform. For a smooth radial field \(q'(0)=0\); all profiles below have
\(\Mmod^{(2)}=O(b^2)\) or faster at the origin, so their shear vanishes
there and their pressure has a finite limit.

For a Gaussian in the local limit,
\begin{equation}
 q(b)=Q_0^2e^{-b^2/(2R^2)},\qquad
 D_*=-\frac{\Lambda_*Q_0^4R^2}{8\pi},
\end{equation}
the analytic benchmark is
\begin{align}
 \Dmod(-\Delta^2)&=\sigma D_*e^{-\Delta^2R^2/4},\label{eq:gaussian-d}\\
 \Dtmod(b)&=\frac{\sigma D_*}{\pi R^2}e^{-b^2/R^2},\\
 \pmodel(b)&=\frac{\Dtmod(b)}{2mR^2}\left(\frac{b^2}{R^2}-1\right),
 \qquad
 \smodel(b)=-\frac{b^2}{mR^4}\Dtmod(b).
 \label{eq:mechanical}
\end{align}
For \(\sigma=+1\), this gives positive core pressure, negative outer
pressure, and positive shear for \(b>0\). The identity
\(\int_0^\infty b\,\dd b\,\pmodel(b)=0\) follows from the boundary
conditions. Stability of the complete hadron requires the full conserved EMT.
Hadronic soliton calculations obtain stress from specified dynamics
\cite{Goeke:2007fp,Cebulla:2007ei}, while Q-ball studies explicitly
distinguish D-term negativity from stability \cite{Mai:2012yc}.
Those dynamical tests and the interpretation questions reviewed in
Ref.~\cite{Lorce:2025oot} concern the complete dynamical system.

To test radial-shape dependence, we also use
\begin{align}
 q_{\rm F}(b)&=Q_0^2\frac{1+\cosh(1/a)}{\cosh[b/(aR)]+\cosh(1/a)},
 \qquad a=0.20,\label{eq:fermi}\\
 q_{\rm P}(b)&=\frac{Q_0^2}{1+(b/R)^4},\\
 q_{\rm L}(b)&=Q_0^2\exp\left[-\frac{(b/R)^2}
 {1+0.35\ln(1+(b/R)^2)}\right].
\end{align}
The symmetrized Fermi profile \(q_{\rm F}\) is even and smooth at the
origin. An unsymmetrized radial Woods--Saxon function has \(q'(0)\ne0\)
and would generate a direction-dependent origin stress and logarithmically
divergent pressure in this model; the symmetrized profile avoids that
singularity. The logarithmically modified profile serves as a separate shape
benchmark.

We set \(Q_0R=mR=\Lambda_*/R^2=1\), \(\sigma=+1\), and \(\ell=0\)
in the local profile comparison. Table~\ref{tab:profiles} reports the direct
moments and finite central pressures. The values scale with the chosen
reference normalization.
For the family \(q_n=Q_0^2/[1+(b/R)^n]\), \(n>1\), the forward moment is
\begin{equation}
 D_{g,n}^{\rm mod}(0)=-\frac{\sigma\Lambda_*Q_0^4R^2}{48\pi}
 n\,\Gamma(2+2/n)\Gamma(2-2/n),
 \label{eq:power-moment}
\end{equation}
which gives \(-1/32\) for the displayed \(n=4\) benchmark. This moment
formula applies under the stated convergence condition; origin smoothness
depends on the chosen exponent \(n\).

\begin{table}[t]
 \centering
 \begin{tabular}{lrr}\toprule
Profile & \(D_g^{\rm mod}(0)\) & \(R^3p_g^{\rm mod}(0)\)\\\midrule
Gaussian & -0.03978874 & 0.00633257 \\
Symmetrized Fermi & -0.03931828 & 0.01149338 \\
Power edge & -0.03125000 & 0.00994718 \\
Log edge & -0.02295474 & 0.01023354 \\
\bottomrule\end{tabular}

 \caption{Local anti-aligned model benchmarks at
 \(Q_0R=mR=\Lambda_*/R^2=1\). Entries are computed directly from the
 infinite-domain coordinate-space integrals, with absolute quadrature
 estimates below \(3\times10^{-11}\) for the forward moments.}
 \label{tab:profiles}
\end{table}

\begin{figure}[t]
 \centering
 \includegraphics[width=\textwidth]{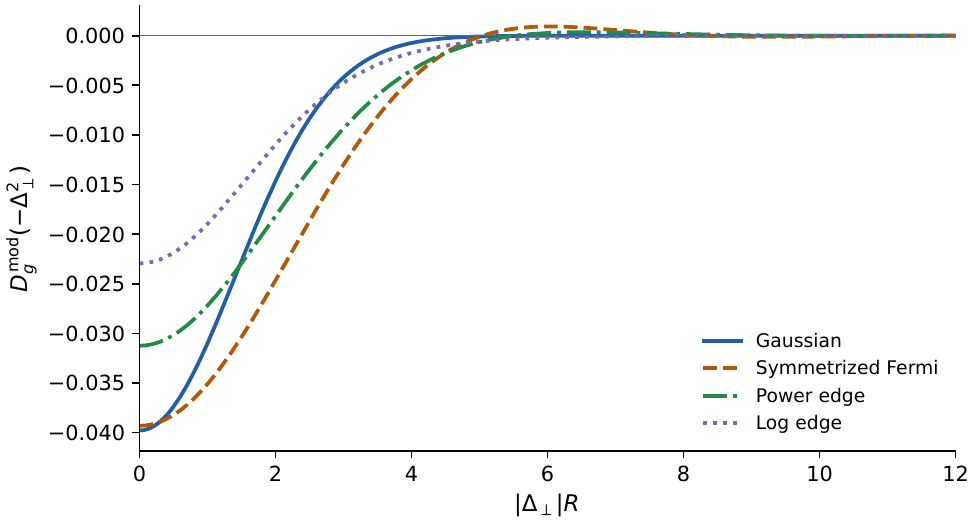}
 \caption{Local anti-aligned model functional for four smooth radial profiles,
 with the normalization of Table~\ref{tab:profiles}. The finite-transfer
 zeroes of the symmetrized Fermi and power-edge curves are genuine features
 of their Bessel projections. The aligned curves are their pointwise negatives
 by linearity.}
 \label{fig:dform}
\end{figure}

\begin{figure}[t]
 \centering
 \includegraphics[width=\textwidth]{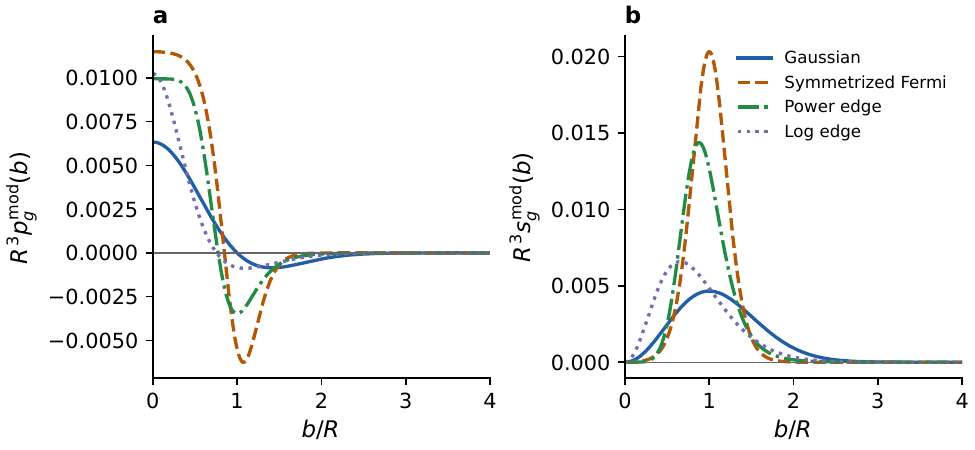}
 \caption{Direct coordinate-space model D-sector stresses for the profiles in
 Figure~\ref{fig:dform}. Panel (a) is the isotropic coefficient and panel (b)
 is the shear coefficient, both with the LF mass rescaling of
 Eq.~\eqref{eq:mass-rescaled-density}. They exclude \(\bar C_g\).
 All profiles have finite central pressure and zero shear at the origin.}
 \label{fig:pressure-shear}
\end{figure}

Figure~\ref{fig:dform} shows the model functional at finite transfer. The
symmetrized Fermi and power-edge profiles have first zeroes near
\(\Delta R=5.04182\) and \(5.51629\), respectively. Their positive
regions show that the sign at finite spacelike transfer can differ from the
forward sign. Figure~\ref{fig:pressure-shear} uses
Eqs.~\eqref{eq:shear-normalization}--\eqref{eq:density-direct} directly.
The shear is nonnegative for every local anti-aligned profile in these
direct coordinate-space curves.


\section{Screened response identities and local-approximation bounds}
\label{sec:finite}
\subsection{A weighted-moment identity for radial profiles}
The sign of the screened forward functional can be established more generally
than by checking the Gaussian pointwise. Let \(f=q_\ell\) and
\(L_b f=f''+f'/b\), so \(q=f-\ell^2L_b f\). Define
\begin{equation}
 I_\ell=\int_0^\infty b^3q'f'\dd b,\qquad
 I_0=\int_0^\infty b^3(q')^2\dd b,
 \qquad D_{g,\ell}^{\rm mod}(0)=-\frac{\sigma\Lambda_*}{8\pi}I_\ell.
\end{equation}
Assume real smooth radial fields, finite integrals below, and a vanishing
boundary term \([b^3f'L_bf-b^2(f')^2]_0^\infty\). Integration by parts gives
\begin{equation}
 \int_0^\infty b^3 f'(L_bf)'\dd b
 =-\int_0^\infty b^3(L_bf)^2\dd b.
 \label{eq:weighted-ibp}
\end{equation}
Writing the three nonnegative norms as
\begin{equation}
 A_f=\int_0^\infty b^3(f')^2\dd b,\quad
 H_f=\int_0^\infty b^3(L_bf)^2\dd b,\quad
 K_f=\int_0^\infty b^3[(L_bf)']^2\dd b,
\end{equation}
we obtain the exact identities
\begin{equation}
 I_\ell=A_f+\ell^2H_f,\qquad
 I_0=A_f+2\ell^2H_f+\ell^4K_f.
 \label{eq:screened-moment-identities}
\end{equation}
For every nontrivial decaying profile in this class,
\begin{equation}
 0<\frac{I_\ell}{I_0}\leq1,\qquad
 I_0-I_\ell=\ell^2H_f+\ell^4K_f\geq0.
 \label{eq:general-screened-bound}
\end{equation}
Thus screening preserves the forward sign selected by \(\sigma\) and reduces
its magnitude relative to the local model. The identity applies to the stated
Helmholtz response for monotonic and nonmonotonic radial profiles, including
those with sign-changing \(q'f'\). It supplies a practical check for a
numerical response solver: the mixed moment equals the positive norms and is
bounded by the corresponding local moment.

A smooth nonmonotonic example illustrates the distinction:
\begin{equation}
 q_{\rm ring}(b)=Q_0^2\frac{b^2}{R^2}e^{-b^2/(2R^2)}.
 \label{eq:ring-profile}
\end{equation}
Here \(I_0=4Q_0^4R^2\). At \(\ell/R=1\), direct quadrature gives
\(I_\ell/I_0=0.46198296\), while
\(R^2q'(1.3R)q_\ell'(1.3R)/Q_0^4=-0.00752606\).
The position-space stress changes sign but its weighted forward moment remains
positive. The code verifies both identities for this profile and the Gaussian
at three response ranges, with the ring derivative also computed independently
in momentum space.

\subsection{Gaussian solution and transfer-window error}
The screened kernel in Eq.~\eqref{eq:screened-kernel} provides a finite-range
extension evaluated with the full momentum dependence of the kernel.
Using \((1+\ell^2k^2)^{-1}=\int_0^\infty\dd u\,e^{-u}e^{-u\ell^2k^2}\),
its action on a Gaussian is
\begin{equation}
 q_\ell(b)=Q_0^2\int_0^\infty\dd u\,e^{-u}
 \frac{R^2}{R^2+2\ell^2u}
 \exp\left[-\frac{b^2}{2(R^2+2\ell^2u)}\right].
 \label{eq:finite-q}
\end{equation}
Its derivative is negative for every \(b>0\). Hence \(q'q_\ell'>0\)
in this explicit background and the anti-aligned forward functional is
negative for every \(\ell\geq0\). This conclusion uses the specified
profile and kernel.

Let \(\eta=\ell/R\) and define
\begin{equation}
 W(u;\eta)=\frac{1+2\eta^2u}{(1+\eta^2u)^3},\qquad
 B(u;\eta)=R^2\frac{1+2\eta^2u}{1+\eta^2u}.
\end{equation}
The Gaussian Bessel integral can then be done before the \(u\) integral:
\begin{equation}
 D_{g,\ell}^{\rm mod}(-\Delta^2)
 =\sigma D_*\int_0^\infty\dd u\,e^{-u}W(u;\eta)
 e^{-\Delta^2B(u;\eta)/4}.
 \label{eq:finite-D}
\end{equation}
This representation is real, finite, and reduces to Eq.~\eqref{eq:gaussian-d}
at \(\ell=0\). For \(\sigma\ne0\), the Gaussian screened functional
keeps its sign at finite transfer, in contrast to the non-Gaussian local
examples. Its transverse density follows by replacing the last
exponential by \(e^{-b^2/B}/(\pi B)\); pressure and shear follow from
Eq.~\eqref{eq:physical-mechanical-def} or the direct stress relations.

\begin{figure}[t]
 \centering
 \includegraphics[width=\textwidth]{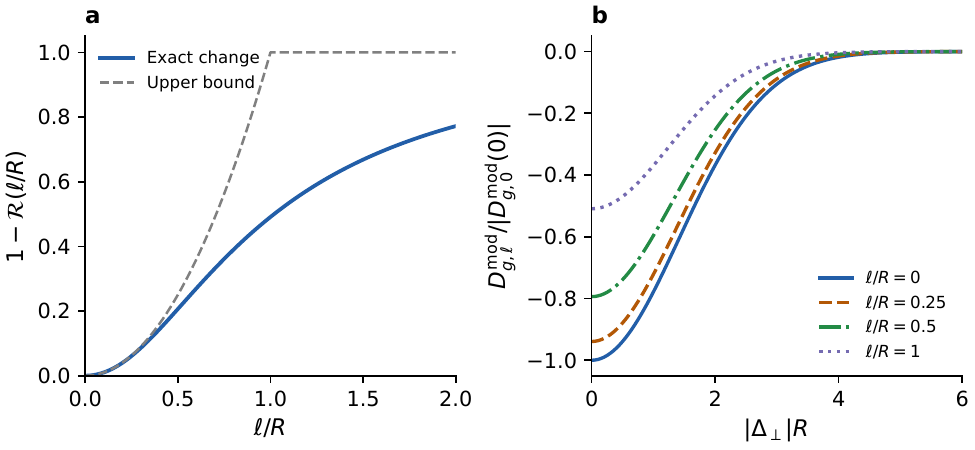}
 \caption{Finite-range Gaussian response with the screened kernel
 \eqref{eq:screened-kernel}. Panel (a) compares the exact forward change
 \(1-\mathcal R\) with the bound \(\min[(\ell/R)^2,1]\).
 Panel (b) shows the anti-aligned model functional normalized by the magnitude
 of its local forward value. Every curve is computed from
 Eq.~\eqref{eq:finite-D}; response range changes its magnitude and
 transfer dependence without changing its sign.}
 \label{fig:finite-range}
\end{figure}

For nonzero \(\sigma\), the forward ratio is
\begin{equation}
 \mathcal R(\eta)\equiv\frac{D_{g,\ell}^{\rm mod}(0)}{D_{g,0}^{\rm mod}(0)}
 =\int_0^\infty\dd u\,e^{-u}\frac{1+2\eta^2u}{(1+\eta^2u)^3}.
 \label{eq:range-ratio}
\end{equation}
For \(z\geq0\),
\(\dd[(1+2z)/(1+z)^3]/\dd z=-(1+4z)/(1+z)^4<0\) and
\(0\leq1-(1+2z)/(1+z)^3\leq z\). Therefore
\begin{equation}
 0<\mathcal R(\eta)\leq1,\qquad
 0\leq1-\mathcal R(\eta)\leq\min(\eta^2,1).
 \label{eq:range-bound}
\end{equation}
The ratio decreases with response range. The bound guarantees that
\(\ell/R\leq0.1\) changes the forward functional by at most one percent
of its local-limit value. Direct quadrature gives
\(\mathcal R(0.1)=0.99001090\),
\(\mathcal R(0.5)=0.79387841\), and
\(\mathcal R(1)=0.50913159\). The comparison is displayed in
Figure~\ref{fig:finite-range}. This quantitative bound is specific to the
Gaussian screened model. Other profile shapes require their own gradient
scales and error estimates.

At finite transfer, an additional bound follows from the full momentum
dependence. With \(a=\Delta^2R^2/4\) and \(z=\eta^2u\), the ratio to the local
Gaussian at the same transfer is
\begin{equation}
 \mathcal R_t(a,\eta)=
 \frac{D_{g,\ell}^{\rm mod}(-\Delta^2)}{D_{g,0}^{\rm mod}(-\Delta^2)}
 =\int_0^\infty\dd u\,e^{-u}\frac{1+2z}{(1+z)^3}
 e^{-az/(1+z)}.
 \label{eq:transfer-ratio}
\end{equation}
Since \(1-We^{-az/(1+z)}\leq(1-W)+az\leq(1+a)z\),
\begin{equation}
 0\leq1-\mathcal R_t(a,\eta)\leq\min[(1+a)\eta^2,1].
 \label{eq:transfer-bound}
\end{equation}
Here \(1-\mathcal R_t\) is the change normalized to the local value,
\(\lvert D_{g,\ell}^{\rm mod}-D_{g,0}^{\rm mod}\rvert/
\lvert D_{g,0}^{\rm mod}\rvert\). A sufficient condition for this ratio
to be at most \(\delta\) throughout \(0\leq\Delta R\leq K\) is
\(\ell/R\leq\sqrt{\delta/(1+K^2/4)}\).
If instead the screened result is the reference for the approximation error,
the relevant ratio is
\begin{equation}
 \frac{\lvert D_{g,\ell}^{\rm mod}-D_{g,0}^{\rm mod}\rvert}
 {\lvert D_{g,\ell}^{\rm mod}\rvert}
 =\frac{1-\mathcal R_t}{\mathcal R_t}
 \leq\frac{B_K}{1-B_K},\qquad
 B_K=(1+K^2/4)\eta^2<1.
 \label{eq:screened-relative-error}
\end{equation}
A sufficient condition for this second ratio to be at most \(\delta\) is
\(\ell/R\leq\sqrt{\delta/[(1+\delta)(1+K^2/4)]}\).
For a one-percent tolerance out to \(\Delta R=6\), conservative rounded
conditions are \(\ell/R\leq0.03162\) with the local denominator and
\(\ell/R\leq0.03146\) with the screened denominator.

\FloatBarrier

\subsection{Shape restrictions and comparison with published pion fits}
\label{sec:lattice-shape}
The Gaussian response family obeys restrictions on its form-factor shape.
Let \(\tau=\Delta^2\),
\(F(\tau)=D_{g,\ell}^{\rm mod}(-\tau)/D_{g,\ell}^{\rm mod}(0)\), and
\(s=-F'(0)>0\), with primes denoting \(\tau\) derivatives. In
Eq.~\eqref{eq:finite-D}, set
\(X=(1+2\eta^2u)/(1+\eta^2u)\) and normalize the weight
\(e^{-u}W(u;\eta)\). Since \(1\leq X<2\),
\(s=R^2\langle X\rangle/4\), and
\begin{equation}
 C\equiv\frac{F''(0)}{[F'(0)]^2}-1
 =\frac{\operatorname{Var}(X)}{\langle X\rangle^2},
 \qquad 0\leq C\leq\frac18.
 \label{eq:lattice-curvature-bound}
\end{equation}
Indeed, \(\langle(X-1)(2-X)\rangle\geq0\) gives
\(\operatorname{Var}(X)\leq(m-1)(2-m)\) for
\(m=\langle X\rangle\); the ratio to \(m^2\) is largest at
\(m=4/3\). Jensen's inequality and the endpoint chord of the convex
exponential also give
\begin{equation}
 e^{-1}\leq F(1/s)\leq\max_{1\leq m\leq2}
 \left[(2-m)e^{-1/m}+(m-1)e^{-2/m}\right]
 =0.389598\ldots.
 \label{eq:lattice-shape-bound}
\end{equation}
These necessary conditions are independent of normalization and of the
chosen response range. They are necessary conditions for membership in the
Gaussian response family.

Published pion gluon fits provide an external illustration.
Reference~\cite{Hackett:2023nkr} gives both monopole and quadratic
\(z\)-expansion fits in the \(\overline{\rm MS}\) scheme at
\(2\,\mathrm{GeV}\), on an ensemble with
\(m_\pi\simeq170\,\mathrm{MeV}\). For the monopole,
\(F(\tau)=1/(1+\tau/\Lambda^2)\), with
\(\Lambda=0.677(65)\,\mathrm{GeV}\). We also reconstruct the second
central curve using its published rounded coefficients
\((\alpha_0,\alpha_1,\alpha_2)=(-0.265,0.682,-0.38)\),
\(D_z(t)=\sum_{k=0}^2\alpha_k z(t)^k\), and
\begin{equation}
 z(t)=\frac{\sqrt{t_c-t}-\sqrt{t_c-t_0}}
 {\sqrt{t_c-t}+\sqrt{t_c-t_0}},\quad
 t_c=4m_\pi^2,\quad
 t_0=t_c\left[1-\sqrt{1+(2\,\mathrm{GeV})^2/t_c}\right].
 \label{eq:pion-z-map}
\end{equation}
Table~\ref{tab:pion-shape} uses \(m_\pi=0.170\,\mathrm{GeV}\).
The normalized \(z\) curve is \(D_z(-\tau)/D_z(0)\).

\begin{table}[htbp]
\centering
\begin{tabular}{lrrrr}
\toprule
Published fit & \(s\,[\mathrm{GeV}^{-2}]\) & \(C\) &
 \(1/s\,[\mathrm{GeV}^2]\) & \(F(1/s)\)\\
\midrule
Monopole & 2.18 & 1.00 & 0.458 & 0.500\\
Quadratic \(z\) expansion & 2.89 & 1.84 & 0.346 & 0.546\\
\bottomrule
\end{tabular}
\caption{Shape diagnostics calculated from the central fits of
Ref.~\cite{Hackett:2023nkr}, Table IV and Eqs.~(64)--(67).
These deterministic reconstructions use the rounded central fit coefficients.
Each last-column entry is evaluated at that fit's own inverse slope.}
\label{tab:pion-shape}
\end{table}

Both central curves violate the Gaussian restrictions: when normalized and
matched to the respective forward slope, every member of the screened
Gaussian family lies at least \(22.1\%\) below the monopole and
\(28.6\%\) below the quadratic \(z\) curve at that curve's inverse
slope. The discrepancy therefore occurs for both central parametrizations.
The forward slopes are extrapolated quantities; statistical comparison with
the underlying correlated lattice results follows in
Section~\ref{sec:lattice-correlated}. These fits describe the full gluon
EMT contribution; the connection of the response model to a small-x
component requires the matching discussed in Section~\ref{sec:operator}.

\subsection{Correlated test with author-supplied pion results}
\label{sec:lattice-correlated}
We now test the shape family against the final combined-irrep pion
estimates associated with Ref.~\cite{Hackett:2023nkr}, supplied by the
authors with their covariance. We use the corrected data file supplied on
9 November 2026: its 24 gluon $D$ entries have indices 1--24, matched to
the corresponding entries of the momentum array. The range is
$0.06888\leq -t\leq1.92209\,\mathrm{GeV}^2$; the forward value is obtained
by extrapolation. The authors confirmed the final combined-irrep convention and inclusion
of all uncertainties carried by those estimates in the supplied covariance,
in response to our request
for the $\overline{\rm MS}$ results at $2\,\mathrm{GeV}$.
The $24\times24$ gluon-$D$ covariance is positive definite, with condition
number 21.29. Its off-diagonal entries are retained throughout.

For $q=-t$, we fit $M_i=\mathcal A F(q_i;s,\eta)$ with free forward
amplitude $\mathcal A$ and slope $s=-F'(0)$. At fixed shape, the amplitude
is profiled analytically using
\begin{equation}
 \widehat{\mathcal A}=\frac{f^TV^{-1}D}{f^TV^{-1}f},\qquad
 \chi^2=(D-M)^TV^{-1}(D-M),\qquad f_i=F(q_i;s,\eta).
 \label{eq:lattice-correlated-objective}
\end{equation}
We evaluate these expressions by Cholesky solves. The range scan uses
$\rho=\eta^2/(1+\eta^2)\in[0,1]$, including the local endpoint and the
limiting normalized shape. In the representation of
Eq.~\eqref{eq:finite-D}, the latter has weight $2x/3$ on $1<x<2$ and
$\langle X\rangle=14/9$. This endpoint completes the normalized shape family
with free amplitude and represents the infinite-range limit. Expanding at fixed slope gives
\begin{equation}
 \log F(q;s,\eta)=-qs+\frac{(qs)^2}{2}\eta^4+O(\eta^6),
 \label{eq:lattice-range-degeneracy}
\end{equation}
for bounded $qs$. Thus the fitted slope absorbs the leading small-range
change, and inference near $\eta=0$ requires a range-profile analysis.

\begin{table}[htbp]
\centering
\begin{tabular}{lrrrrr}
\toprule
Transfer window & Points & Local & Screened & Monopole & Quadratic $z$\\
\midrule
All supplied $D_g$ points & 24 & 27.820 & 27.501 & 22.404 & 22.548\\
$q\leq1\,\mathrm{GeV}^2$ & 13 & 14.799 & 14.681 & 13.311 & 13.671\\
$q\leq0.5\,\mathrm{GeV}^2$ & 7 & 7.456 & 7.456 & 8.086 & 6.650\\
\bottomrule
\end{tabular}
\caption{Minimum correlated $\chi^2$ values. The screened fit includes its endpoint limits.
The full and $q\leq1$ fits minimize at $\rho=1$; the lowest-transfer fit
minimizes at $\rho=0$. The two restricted windows were specified before
fitting the corrected data and are sensitivity checks. The point counts
denote correlated momentum bins from the same gauge ensemble.}
\label{tab:pion-correlated}
\end{table}

The full-window monopole refit gives $D_g(0)=-0.5792$ and
$\Lambda=0.6842\,\mathrm{GeV}$. For the quadratic-$z$ fit we obtain
$\alpha_0=-0.2609$, $\alpha_1=0.6638$, and $\alpha_2=-0.3537$.
Here we use Eq.~\eqref{eq:pion-z-map} with $m_\pi=0.170\,\mathrm{GeV}$.
These values are close to the published summaries, with differences in the
rounded central values and $\chi^2/\mathrm{d.o.f.}$; the source of those
differences remains to be identified. Our numbers are independent refits of
the corrected supplied estimates. Changing the mass in the $z$ map to
$0.169\,\mathrm{GeV}$ changes its minimum $\chi^2$ by 0.0024; the Gaussian
response fits use a separate parametrization.

\begin{figure}[htbp]
\centering
\includegraphics[width=\linewidth]{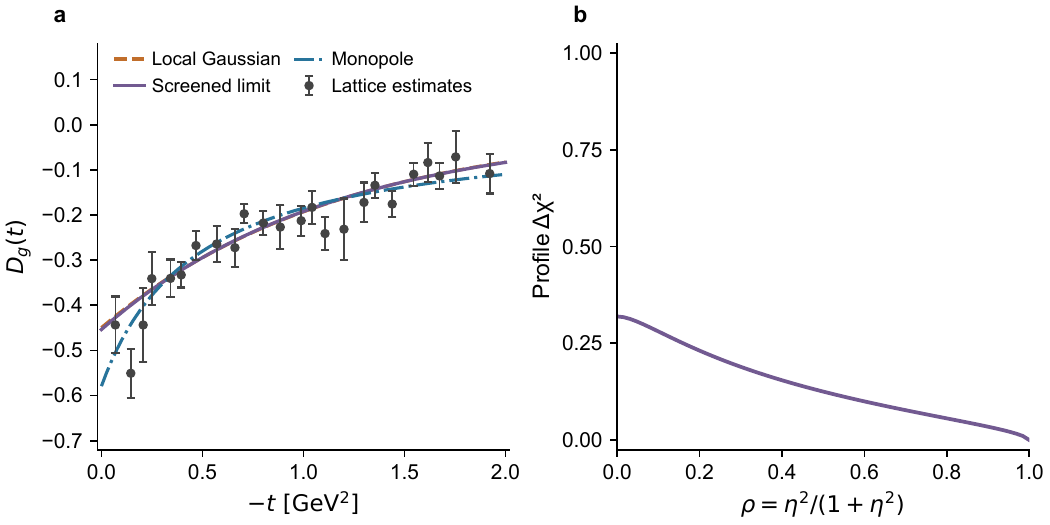}
\caption{Correlated comparison using the corrected pion estimates.
(a) The 24 central values with marginal one-standard-deviation error bars;
all fits use the complete gluon-$D$ covariance. Lines show fitted central
curves. The screened line is the normalized shape limit
$\rho=1$. (b) Profile $\chi^2(\rho)-\chi^2_{\min}$ with amplitude and slope
refitted at every range. The vertical scale displays profile differences;
their endpoint calibration is given in the text.}
\label{fig:pion-correlated}
\end{figure}

The full-window local and screened amplitudes are $-0.4497$ and $-0.4537$,
respectively, and their slopes are $0.8476$ and
$0.8722\,\mathrm{GeV}^{-2}$. The range profile spans only
$\Delta\chi^2=0.3195$, with its minimum at the infinite-range shape limit.
Figure~\ref{fig:pion-correlated} shows the fitted curves and the profile;
the statistical interpretation follows from the simulation below.

To assess the two distinct questions of fit quality and range preference,
we generate correlated Gaussian pseudo-data with the supplied covariance
held fixed and refit each draw. For goodness of fit, 4096 draws at each
model's fitted mean give tail fractions 0.185 for the local Gaussian and
0.448 for the monopole. The respective 95\% binomial Monte Carlo intervals
are $[0.173,0.197]$ and $[0.432,0.463]$. At the 5\% level, both models
are compatible with these data in this goodness-of-fit test under the
fixed-covariance Gaussian approximation.

For the improvement statistic
$T=\chi^2_{\rm local}-\min_{\rho\in[0,1]}\chi^2(\rho)$,
25 of 2048 pseudo-data sets drawn at the fitted local null exceed the
observed $T=0.3195$. With the plus-one convention the estimated tail
probability is 0.0127; its binomial Monte Carlo interval is
$[0.0083,0.0180]$. The bounded alternative has a nonstandard, narrow null
distribution; its calibration is therefore essential to interpret $T$.
The calculation draws pseudo-data from the fitted local model with the
supplied covariance held fixed. Accounting for uncertainty in that covariance
requires additional information about its estimation.
To test nuisance sensitivity, an independent simulation with 2048 draws per
case and common random numbers gives tail fractions 0.0205 at the fitted
local slope and 0.0556 at $s=1.03\,\mathrm{GeV}^{-2}$. The latter slope,
with amplitude profiled, is $3.409$ in $\chi^2$ above the local minimum.
The respective 95\% binomial Monte Carlo intervals are $[0.0148,0.0270]$
and $[0.0461,0.0659]$. The second interval overlaps 0.05. These two
local-null choices therefore illustrate the sensitivity of the calibration;
uniform coverage across nuisance parameters remains to be established.

The full-window best fit lies at the infinite-range shape endpoint, while the
lowest-transfer window favors the local endpoint. Forward extrapolation also
depends on the model: the local, monopole, and quadratic-$z$ refits give
$D_g(0)=-0.450,-0.579,-0.601$, respectively. Within the local and monopole
models separately, the same simulations give 95\% percentile intervals
$[-0.516,-0.386]$ and $[-0.717,-0.468]$. These conditional intervals quantify
within-model uncertainty; uncertainty from choosing a shape family requires
a separate treatment. Applying the small-range error bound requires
an independent constraint establishing $\eta\ll1$. The comparison tests
the phenomenological shape family against the full gluon form factor; its
small-x component requires the QCD matching discussed in
Section~\ref{sec:operator}.

\FloatBarrier

\section{A color-singlet Coulombic stress benchmark}
\label{sec:quarkonium}
We now compare the local and filtered mixed fields in one normalized target
model. Consider a weakly coupled, spin-zero, equal-mass heavy quark--antiquark
singlet with the leading Coulomb potential. Its normalized relative wave
function and Bohr radius are
\begin{equation}
 \psi_{1S}(\bm r)=\frac{e^{-r/a_0}}{\sqrt{\pi a_0^3}},\qquad
 a_0=\frac{2}{m_Q C_F\alpha_s},\qquad C_F=\frac{N_c^2-1}{2N_c},
 \label{eq:quarkonium-wavefunction}
\end{equation}
with a fixed weak coupling \(\alpha_s=g^2/(4\pi)\) and \(M\simeq2m_Q\). A Coulombic regime requires
the inverse bound-state scale to be perturbative. We use it as a
leading-potential benchmark. Effective theories
of such weakly coupled heavy pairs have a broader operator and matching
structure \cite{Brambilla:1999xf}.

In a fixed Coulomb gauge, let \(\rho^a(\bm k)\) denote the Fourier-transformed
leading color density of the two particles. Summing over the color index in
the normalized singlet gives the non-diagonal covariance
\begin{align}
 \sum_a\langle\rho^a(\bm k)\rho^a(\bm k')^\dagger\rangle
 &=2g^2C_F h_s(k)h_s(k')
 \left[P\!\left(\frac{\bm k-\bm k'}2\right)
      -P\!\left(\frac{\bm k+\bm k'}2\right)\right],
 \label{eq:quarkonium-covariance}\\
 P(\bm q)&=\left(1+\frac{a_0^2\bm q^2}{4}\right)^{-2},
 \qquad h_s(k)=e^{-s^2k^2/2}.\nonumber
\end{align}
The second term is the quark--antiquark color interference; the covariance
vanishes when either momentum is zero. The Gaussian factor is an ultraviolet
regulator in this fixed-gauge calculation.
The leading electric potential field follows from Gauss's law,
\(E_i^a(\bm k)=-\ii k_i\rho^a(\bm k)/\bm k^2\).
The stress calculation retains the full two-field expectation value.

The leading electric contribution to the static Breit-frame stress uses the
same spin-zero normalization as Eq.~\eqref{eq:state-normalization}. In a
frame with transverse transfer and \(E_\Delta=\sqrt{M^2+\Delta^2/4}\),
the two-dimensional STF projector extracts its contribution through
\begin{equation}
 \frac{\langle p'|T^{ij}_{g,\rm STF}|p\rangle_{\rm pot}}{2E_\Delta}
 =\frac{\Delta^i\Delta^j-\delta^{ij}\Delta^2/2}{4E_\Delta}
 D_{\rm pot}(t).
 \label{eq:quarkonium-breit-projector}
\end{equation}
The equal-time potential-field calculation is evaluated at leading nonrelativistic
order and forward transfer. In the point-source limit it reproduces the
known two-body static-vector result \cite{Freese:2026hjy}, here with the
singlet color factor,
\begin{equation}
 \frac{D_{\rm pot}(0)}{M}=\frac{7g^2C_Fa_0}{40\pi},\qquad
 D_{\rm pot}(0)\simeq\frac{14}{5}
 \quad\text{for Eq.~\eqref{eq:quarkonium-wavefunction} and } M\simeq2m_Q.
 \label{eq:quarkonium-forward}
\end{equation}
The positive result is the potential-field contribution in this benchmark.
Relating it to the renormalized \(D_g(0;\mu)\) of a physical quarkonium
requires hard matching, quark--gluon operator mixing, dynamical transverse
gluons, and relativistic corrections. An \(x\)-resolved small-\(x\) form
factor requires an additional longitudinal construction.

For a frozen source at this static potential order, a longitudinal boost of factor \(\zeta\)
gives \(F^{i-}=\zeta^{-2}F^{i+}\). The co-moving field is therefore local
at this order. To quantify the
effect of imposing one, we deliberately filter one mixed field by
\(K_\ell(k_\perp)=(1+\ell^2k_\perp^2)^{-1}\), while holding the target
and the other field fixed. We denote the corresponding transverse STF
functional by \(\mathcal D_\ell^{(s)}\) and normalize it to its unfiltered
value,
\begin{equation}
 \mathcal R_Q(h,r_s)=\frac{\mathcal D_\ell^{(s)}(0)}
 {\mathcal D_0^{(s)}(0)},\qquad h=\frac\ell{a_0},\quad r_s=\frac{s}{a_0}.
 \label{eq:quarkonium-ratio}
\end{equation}
A transverse Wilson line gives this particular filter a gauge-covariant
completion. At fixed \(x^\pm\), let \(U_\perp(x,z)\) be a straight Wilson
line from \(\bm z_\perp\) to \(\bm x_\perp\), and set
\begin{equation}
 \mathcal F^{i-}_\ell(x)=\int\dd^2z_\perp\,
 \frac{K_0(|\bm x_\perp-\bm z_\perp|/\ell)}{2\pi\ell^2}\,
 U_\perp(x,z)F^{i-}(z)U_\perp(z,x).
 \label{eq:quarkonium-link}
\end{equation}
It transforms at the point \(x\), so replacing \(F^{i-}(x)\) by
\(\mathcal F^{i-}_\ell(x)\) inside the color trace of
Eq.~\eqref{eq:local-stress} gives a gauge-invariant bilocal construction.
In the leading potential background \(A_\perp=0\), the line is unity and
Eq.~\eqref{eq:quarkonium-link} reduces to the filter used in the calculation.
The path and Gaussian source regulator are model choices. The filtered
field is not a solution for the same current; process-dependent and
renormalized QCD matching require an independent calculation.
Appendix~\ref{app:quarkonium} gives the explicit color-state and
Schwinger-parameter calculation.

For point sources, the same-target comparison gives
\begin{equation}
 1-\mathcal R_Q(h,0)=\frac{5\pi}{14}h-\frac{88}{147}h^2+o(h^2).
 \label{eq:quarkonium-linear}
\end{equation}
At \(h=0.02\) the calculated change relative to the unfiltered value is
\(2.220\%\). The linear behavior reflects the point-source singularities
of the nonfactorized two-body field. Equations~\eqref{eq:range-bound}--
\eqref{eq:transfer-bound} apply to the smooth single radial Gaussian.

\begin{figure}[t]
 \centering
 \includegraphics[width=\textwidth]{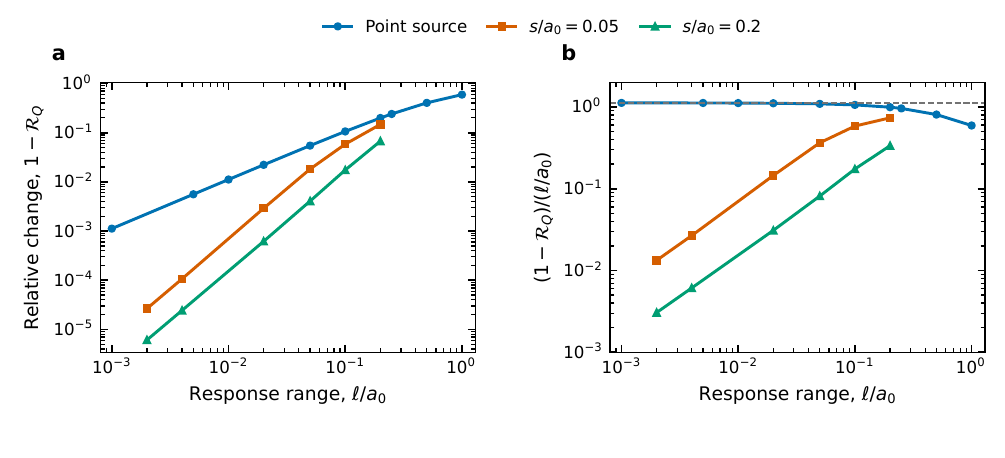}
 \caption{Finite-range replacement in the same leading-potential target.
 (a) Change relative to the unfiltered source at the same ultraviolet
 regulator. (b) Division by \(h=\ell/a_0\) exposes the linear point-source
 limit \(5\pi/14\) (dashed line); at fixed \(s>0\) the quotient tends to
 zero. Points are calculated values; lines connect them. The filtered
 functional belongs to the specified leading-potential benchmark.}
 \label{fig:quarkonium}
\end{figure}

The point-source result is sensitive to ultraviolet modes. With
the same target state and a fixed finite \(s\), the leading small-\(h\)
change instead scales as \(h^2\). Table~\ref{tab:quarkonium} and
Figure~\ref{fig:quarkonium} display the regulator dependence. Taking
\(s\to0\) before \(h\to0\) and taking the small-\(h\) derivative at fixed
\(s\) give different limits. Since the point-source limit can probe momenta
above the domain of the potential approximation, a physical interpretation
requires a gauge-consistent ultraviolet prescription and operator matching.

\begin{table}[t]
 \centering
 \begin{tabular}{lrrr}\toprule
 \(s/a_0\) & \(\mathcal R_Q(0.02,r_s)\) & \(1-\mathcal R_Q\) &
 \(h\) at one percent \\\midrule
 \(0\) & 0.977800 & 2.220\% & 0.008955 \\
 \(0.05\) & 0.997095 & 0.290\% & 0.036567 \\
 \(0.20\) & 0.999381 & 0.0619\% & 0.076712 \\\bottomrule
 \end{tabular}
 \caption{Same Coulombic singlet target with different fixed-gauge source
 regulators. Errors are normalized to the unfiltered result at the same
 \(s\); the last column gives the numerical range at a one-percent change
 within this regulated benchmark.}
 \label{tab:quarkonium}
\end{table}
\FloatBarrier

\section{Discussion and conclusions}
\label{sec:discussion}
The zero-skewness spin-zero D-term projector isolates a mixed transverse
field-strength product and identifies the stress information needed beyond
the leading-plus-field background. The standard EMT projector, renormalized
GPD moment, and LF mechanical relations fix the physical normalization.
For finite recoil at fixed transverse transfer, the extraction includes the
\(A_g\)-slope correction in Eq.~\eqref{eq:recoil-corrected-D}. The standard
ERBL freedom identifies an independent contribution to the D-term that
requires finite-skewness input.

For the explicitly sourced commuting model, the weighted radial identity
fixes the screened forward sign and bounds its magnitude by the local value.
It provides a diagnostic for general admissible radial profiles, including
ones with sign-changing position-space stress. The Gaussian response also
admits a positive-weight representation and explicit error bounds for a local
approximation over a chosen transfer window. Direct coordinate-space
mechanical relations remove inverse-transform artifacts and give regular
benchmarks for four smooth profiles. These results concern the specified
screened response kernel; its commuting color structure and independent
normalization remain model assumptions. The curvature and finite-transfer
shape restrictions provide an additional test of this family. Both published
pion central fits considered in Section~\ref{sec:lattice-shape} lie outside
these restrictions when their forward slopes are matched. In the correlated
reanalysis of Section~\ref{sec:lattice-correlated}, the local Gaussian gives
an acceptable overall fit under a fixed-covariance Gaussian approximation.
The full-window optimum lies at the limiting screened shape, while the
lowest-transfer window favors the local endpoint. The range-test calibration
depends on the assumed local-null slope. The data therefore test the model shapes, while the fitted response length
remains unresolved and the forward extrapolation depends on the shape family.

The Coulombic singlet benchmark adds a same-target comparison with a
normalized two-body wave function. The leading static potential fixes a
co-moving mixed-field ratio, while the deliberately imposed finite-range
replacement is sensitive to ultraviolet regularization. Its linear
point-source behavior contrasts with the smooth radial Gaussian bound and
illustrates the role of short-distance structure in the response.

The sign variation with \(\sigma\) concerns externally sourced configurations.
An assessment of hadron stability requires the source stress, quark
contributions, and the conserved total EMT. To connect the decorated-line
construction to a physical gluon D-term, the next ingredients are its
longitudinal distribution and moment, rapidity regularization, operator
mixing, and matching to the renormalized local EMT. A specified
noncommuting target ensemble would then allow the model sign conditions to be
tested against a physical matrix element.

\acknowledgments
We thank Phiala Shanahan for providing the pion form-factor estimates and
covariance, and for clarifying the corrected data indexing.
This work is supported in part by the National Natural Science Foundation of
China (NSFC) under Grant No. 1234710148, and in part by the China Postdoctoral
Science Foundation under Grant No. 2023M742098.
\href{https://openai.com/codex/}{OpenAI Codex} was used to assist with manuscript restructuring, algebraic
cross-checks, reference verification, and numerical implementation.
The analytic and numerical consistency checks are described in this paper
and its accompanying reproduction material. Responsibility for the scientific
content and submission remains with the author.

\paragraph{Data and code availability.}
All model inputs and analytic expressions are specified in this paper.
The reproduction script, numerical tables, and validation results are supplied
in the accompanying supplementary source material, including the Coulombic
benchmark. The pion study includes both a reconstruction of published central
curves and an independent correlated refit of corrected form-factor estimates
provided by the authors of Ref.~\cite{Hackett:2023nkr}. The analysis scripts,
fit summaries, and fixed-seed parametric-simulation specifications are retained
with the reproduction materials. The privately supplied numerical attachment
is not redistributed here; requests for the underlying estimates and
covariance should be directed to the original authors. A separate script
checks the GPD normalization, ERBL moments, and on-shell recoil subtraction;
that illustration uses published central A and D fits. Gaussian pseudo-data
used to calibrate the fits are generated from the supplied covariance; the
underlying lattice results are those of the cited work.

\appendix
\section{Current supporting the commuting background}
\label{app:source}
Write \(a(z^-)=\kappa h(z^-)\) and \(c=\sigma\epsilon R_0\).
Using contravariant light-cone fields in Eq.~\eqref{eq:background}, the
nonzero independent strengths are
\begin{equation}
 F^{i+}=-a\,\partial_iq\,H,\qquad
 F^{i-}=c a\,\partial_iq_\ell\,H,\qquad
 F^{+-}=-c a' q_\ell H.
\end{equation}
All fields are independent of \(x^+\). The resulting current is
\begin{equation}
 J^+=(c a''q_\ell-a\nabla_b^2q)H,\qquad
 J^-=c a\nabla_b^2q_\ell H,\qquad
 J^i=-c a'\partial_iq_\ell H.
\end{equation}
It obeys \(\partial_+J^++\partial_-J^-+\partial_iJ^i=0\) directly.
All commutators with the potentials vanish, so ordinary and covariant
conservation coincide here. The additional \(F^{+-}\) contributes to
other EMT components, while its contribution to the STF projection in
Eq.~\eqref{eq:local-stress} vanishes. Calculations of the total stress or
source energy retain this field. The current is specified externally.

For clarity, the derivative tangent to a Wilson line parametrized by
\(z^-\) is \(D_-=D^+\), containing \(A^+\).
The operator \(D^-=D_+\) differentiates \(x^+\); its integral along
\(z^-\) requires a separate evaluation and is not generally an endpoint term.

\section{Transform identities and convergence}
\label{app:transforms}
Choosing the transverse transfer along the first axis gives
\begin{equation}
 Q^{ij}\Mmod^{ij}=\tfrac12\Delta^2\cos(2\phi)\Mmod^{(2)}(b).
\end{equation}
The angular identity
\begin{equation}
 \int_0^{2\pi}\dd\phi\,e^{-\ii\Delta b\cos\phi}\cos2\phi
 =-2\pi J_2(\Delta b)
\end{equation}
then proves Eq.~\eqref{eq:bessel}. The limit \(J_2(z)/z^2\to1/8\)
gives the forward formula whenever its absolute weighted moment converges,
\begin{equation}
 \int_0^\infty b^3\big|\Mmod^{(2)}(b)\big|\dd b<\infty.
\end{equation}
The same proof applies to the physical Fourier density without a model label.

Writing \(f=\Dtmod\), the STF Fourier relation gives
\begin{equation}
 f''-\frac{f'}b=-\frac{\Mmod^{(2)}}{2\pi^2},\qquad
 \frac{f'(b)}b=\frac1{2\pi^2}\int_b^\infty\frac{\dd u}{u}\Mmod^{(2)}(u).
\end{equation}
Integrating once more with the decaying boundary condition gives
Eq.~\eqref{eq:density-direct}. Substituting into
Eq.~\eqref{eq:physical-mechanical-def} proves the pressure and shear
relations. Smooth local profiles obey \(\Mmod^{(2)}=O(b^2)\) at the
origin; the power-edge benchmark decays as \(b^{-10}\) at infinity and
the other profiles decay faster. All required integrals converge.
For these profiles the pressure integral vanishes by exchanging convergent
integrals, or equivalently by the boundary term \([bf']_0^\infty\).

The numerical calculation uses analytic profile derivatives and adaptive
quadrature on the infinite radial domain. The stable quantity
\(J_2(\Delta b)/(\Delta b)^2\) is evaluated with its series near zero.
There are 241 displayed transfer points on \(0\leq\Delta R\leq12\)
and 601 radial points on \(0\leq b/R\leq6\); the integrals use the full
radial domain. The forward moments are checked against the Gaussian
and power-family analytic results, changes of quadrature tolerance, and
finite-radius tail integrals. Local stresses are evaluated directly, including
at the origin. For the screened Gaussian, an independent momentum-space
integral of the smoothed profile derivative and a subsequent radial stress
projection check Eq.~\eqref{eq:finite-D}. The accompanying validation
report records 71 base checks and 41 additional screened-moment and
transfer-bound checks, together with quadrature estimates and tail tests. Quadrature
estimates and tolerance comparisons quantify numerical accuracy; model
assumptions are specified separately.

For the nonmonotonic ring test, the independent momentum-space derivative is
\begin{equation}
 q_{\ell,\rm ring}'(b)=-Q_0^2R^2\int_0^\infty\dd k\,
 k^2J_1(kb)\frac{(2-R^2k^2)e^{-R^2k^2/2}}{1+\ell^2k^2}.
\end{equation}
It is compared with the derivative obtained by heat evolution of
Eq.~\eqref{eq:ring-profile}. The three norms in
Eq.~\eqref{eq:screened-moment-identities} and the mixed moment are integrated
separately. The additional report also records the finite-transfer bound
at \(\Delta R=0,1,4,6\) and \(\ell/R=0.1,0.5,1\).

\section{Coulombic response calculation}
\label{app:quarkonium}
The relative probability in Eq.~\eqref{eq:quarkonium-wavefunction} has
Fourier transform \(P(\bm q)=(1+a_0^2\bm q^2/4)^{-2}\). The equal-mass
center coordinate puts the particles at \(\pm\bm r/2\). The singlet
identities
\begin{equation}
 \sum_a\langle T_1^aT_1^a\rangle=C_F,\qquad
 \sum_a\langle T_1^aT_2^a\rangle=-C_F
\end{equation}
give
Eq.~\eqref{eq:quarkonium-covariance}. The point-source result
\eqref{eq:quarkonium-forward} follows independently from the two-body
static-vector forward expression of Ref.~\cite{Freese:2026hjy} with equal
masses and opposite color charges, followed by the singlet color contraction.

For the finite-range comparison, represent the Helmholtz factor by
\begin{equation}
 K_\ell(k_\perp)=\int_0^\infty\dd u\,e^{-u-u\ell^2k_\perp^2}.
\end{equation}
The 1S form factor is a Gaussian scale mixture with weight \(ve^{-v}\):
the variance of either particle's position is \(B^2=a_0^2v/8\).
Combining the two Coulomb denominators using Schwinger parameters yields
\begin{align}
 \frac{\mathcal D_\ell^{(s)}(0)}{2g^2C_F M}
 &=\int_0^\infty\dd v\,v e^{-v}\,
 I_\ell\!\left(\frac{a_0^2v}{4},s\right),\label{eq:quarkonium-I}\\
 I_\ell(L,s)&=\frac1{8\pi^{3/2}}\int_0^\infty\dd u\,e^{-u}
 \int_{s^2}^\infty\dd p\,p\chi_s(p)
 \left[\mathcal K_0(p,\ell^2u)-\mathcal K_L(p,\ell^2u)\right],\\
 \mathcal K_L(p,\alpha)&=\frac{
 1-[p^2\chi_s(p)^2/3+\alpha^2]/(p+L+\alpha)^2}
 {(p+L+\alpha)\sqrt{p+L}},\qquad
 \chi_s(p)=1-\frac{s^2}{p}.
 \label{eq:quarkonium-K}
\end{align}
The two terms in brackets correspond to the self and interference
contributions, combined before the forward integral. At \(\ell=0\),
\begin{equation}
 I_0(L,s)=\frac{7}{30\pi^{3/2}}\left(\sqrt{s^2+L}-s\right).
 \label{eq:quarkonium-I0}
\end{equation}

For \(s=0\), define
\begin{align}
 J(\beta)&=\int_0^1\dd x\left[
 \frac{4x^4(x^2+3\beta)}{3(x^2+\beta)^3}
 +\frac{(4\beta+2)x^2-\tfrac23x^4}{(1+\beta x^2)^3}\right],
 & J(0)&=\frac{28}{15},\label{eq:quarkonium-J}
\end{align}
and \(\beta(h,y)=4h^2y^2/(1-y^2)\). Then
\begin{equation}
 \mathcal R_Q(h,0)=1-\frac{4h}{3J(0)B(3/2,2)}
 \int_0^1\dd y\,2y^2(1-y^2)
 \frac{J(0)-J(\beta)}{\sqrt\beta}.
 \label{eq:quarkonium-beta-prime}
\end{equation}
Here \(B\) is the Euler beta function, and the integrand is understood by
continuity at its endpoints.
Expansion of Eq.~\eqref{eq:quarkonium-J} gives
\(J(\beta)=28/15-(\pi/2)\sqrt\beta+(44/105)\beta+O(\beta^{3/2})\),
which leads to Eq.~\eqref{eq:quarkonium-linear}. At fixed \(s>0\), the
lower limit \(p=s^2\) permits expansion in \(\ell^2\); the coefficients
are regulator dependent.

The supplied script checks the SU(3) singlet identities and covariance
positivity, compares the analytic 1S characteristic function with direct
radial integration, and checks Eq.~\eqref{eq:quarkonium-I0} against an
independent Gaussian Schwinger integral. It also compares finite-transfer
three-dimensional momentum quadrature to the Schwinger representation
before taking the forward limit, and tests the point-source formula against
an independent two-variable Gaussian-mixture quadrature. The finite-\(s\)
table is checked by increasing all quadrature orders. These checks address
the numerical accuracy of the specified leading-potential model.

\bibliographystyle{JHEP}
\bibliography{references}
\end{document}